\documentclass[graybox]{svmult}

\usepackage{natbib}
\usepackage{mathptmx}       
\usepackage{helvet}         
\usepackage{courier}        
\usepackage{type1cm}        
\usepackage{makeidx}         
\usepackage{graphicx}        
\usepackage{multicol}        
\usepackage[bottom]{footmisc}

\makeindex             
\newcommand{\apj}{ApJ}
\newcommand{\apjl}{ApJ}
\newcommand{\apjs}{ApJS}
\newcommand{\aap}{A\&A}

\newcommand{\mnras}{MNRAS}

\newcommand{\physrep}{Physics Reports}

\newcommand{\araa}{ARAA}

\newcommand{\pasa}{PASA}

\newcommand{\procspie}{PROCSPIE}
\newcommand{\hs}{\hspace{1mm}}
\def\gsim{~\rlap{$>$}{\lower 1.0ex\hbox{$\sim$}}}
\newcommand{\taud}{\tau_{\rm D}}
\newcommand{\tauHII}{\tau_{\rm HII}}
\def\lsim{~\rlap{$<$}{\lower 1.0ex\hbox{$\sim$}}}

\begin{document}

\title*{Constraining Reionization with Ly$\alpha$ Emitting Galaxies}
\author{Mark Dijkstra}
\institute{Institute for Theoretical Astrophysics,  University of Oslo, Postboks 1029, 0315 Oslo \email{mark.dijkstra@astro.uio.no}}
%
%




\maketitle
\vspace{-30mm}

\abstract*{Neutral diffuse intergalactic gas that existed during the Epoch of Reionization (EoR) suppresses Ly$\alpha$ flux emitted by background galaxies. In this chapter I summarise the increasing observational support for the claim that Ly$\alpha$ photons emitted by galaxies at $z>6$ are suppressed by intervening HI gas. I describe key physical processes that affect Ly$\alpha$ transfer during the EoR. I argue that in spite of the uncertainties associated with this complex multiscale problem, the data on Ly$\alpha$ emitting galaxies at $z=0-6$ strongly suggests that the observed reduction in Ly$\alpha$ flux from galaxies at $z>6$ is due to additional intervening HI gas. The main question is what fraction of this additional HI gas is in the diffuse neutral IGM. I summarise how future surveys on existing and incoming instruments are expected to reduce existing observational uncertainties enormously. With these improved data we will likely be able to nail down reionization with Ly$\alpha$ emitting galaxies.}
\abstract{Neutral diffuse intergalactic gas that existed during the Epoch of Reionization (EoR) suppresses Ly$\alpha$ flux emitted by background galaxies. In this chapter I summarise the increasing observational support for the claim that Ly$\alpha$ photons emitted by galaxies at $z>6$ are suppressed by intervening HI gas. I describe key physical processes that affect Ly$\alpha$ transfer during the EoR. I argue that in spite of the uncertainties associated with this complex multiscale problem, the data on Ly$\alpha$ emitting galaxies at $z=0-6$ strongly suggests that the observed reduction in Ly$\alpha$ flux from galaxies at $z>6$ is due to additional intervening HI gas. The main question is what fraction of this additional HI gas is in the diffuse neutral IGM. Current models favor a volume filling factor of neutral gas of $x_{\rm HI}>0.4$ at $z~7$. I summarise how future surveys on existing and incoming instruments are expected to reduce existing observational uncertainties enormously. With these improved data we will likely be able to nail down reionization with Ly$\alpha$ emitting galaxies.}
\vspace{-5mm}
\section{Introduction}
\label{sec:intro}
\vspace{-3mm}
Diffuse, neutral intergalactic gas that existed during the Epoch of Reionization (EoR) was opaque to Ly$\alpha$ emission emitted by background galaxies. The spatial distribution of diffuse neutral intergalactic gas may thus be manifest in the distribution of Ly$\alpha$ emitting galaxies during the EoR \citep[e.g.][]{Haiman99,MR04}. A sudden reduction in the amount of Ly$\alpha$ flux from galaxies at $z>6$ has been detected (see \S~\ref{sec:obs}), which is explained naturally by the emergence of neutral intergalactic gas at these epochs.

In this chapter, I summarise observational support for the claim that Ly$\alpha$ flux is suppressed significantly more than expected from lower-redshift observations in \S~\ref{sec:obs}. I then describe key physical processes that regulate the transfer of Ly$\alpha$ photons through the interstellar medium (ISM) and intergalactic medium (IGM) in \S~\ref{sec:key}. Our current understanding of these processes suggests that the observed reduction in Ly$\alpha$ flux from galaxies at $z>6$ at face value requires a large (volume-averaged) neutral fraction of $x_{\rm HI} > 0.4$ at $z=7$, although the observational uncertainties are still large (see \S~\ref{sec:interpret}). I then summarise in \S~\ref{sec:future} how Ly$\alpha$ emitting galaxies detected in future surveys can be used to place robust constraints on the ionisation state of the IGM during the EoR, and will focus on ({\it i}) the redshift evolution of Ly$\alpha$ luminosity functions, ({\it ii}) observed clustering of Ly$\alpha$-selected galaxies, and ({\it iii}) the cross-correlation between LAEs and the 21-cm signature from neutral intergalactic gas. I summarise in \S~\ref{sec:conc}.

Throughout this chapter, I refer to Ly$\alpha$-selected galaxies as `{\it Ly$\alpha$ emitters (LAEs)}'. The term `LAE'  is also used to refer to all star forming galaxies with a Ly$\alpha$ emission line with a rest frame equivalent width (EW) that exceeds EW$>20$ \AA. In practise, this distinction makes little difference, although I caution that some narrow-band surveys for LAEs use color-criteria which correspond to LAEs having a rest-frame EW as high as EW$\gsim 64$ \AA. I use the term `{\it Ly$\alpha$ emitting galaxies}' to refer to galaxies with `strong' Ly$\alpha$ emission irrespective of how they were selected: Ly$\alpha$ emitting galaxies thus include LAEs and drop-out galaxies with strong Ly$\alpha$ emission.

\vspace{-5mm}
\section{Observational Support for a Reduced Ly$\alpha$ Flux from Galaxies at $z>6$}
\label{sec:obs}
\vspace{-4mm}
There are two complementary lines of evidence that support a sudden reduction in the Ly$\alpha$ flux from galaxies at $z>6$:\\

\begin{figure}[tb]
\centerline{\hspace{0mm}\includegraphics[scale=.45]{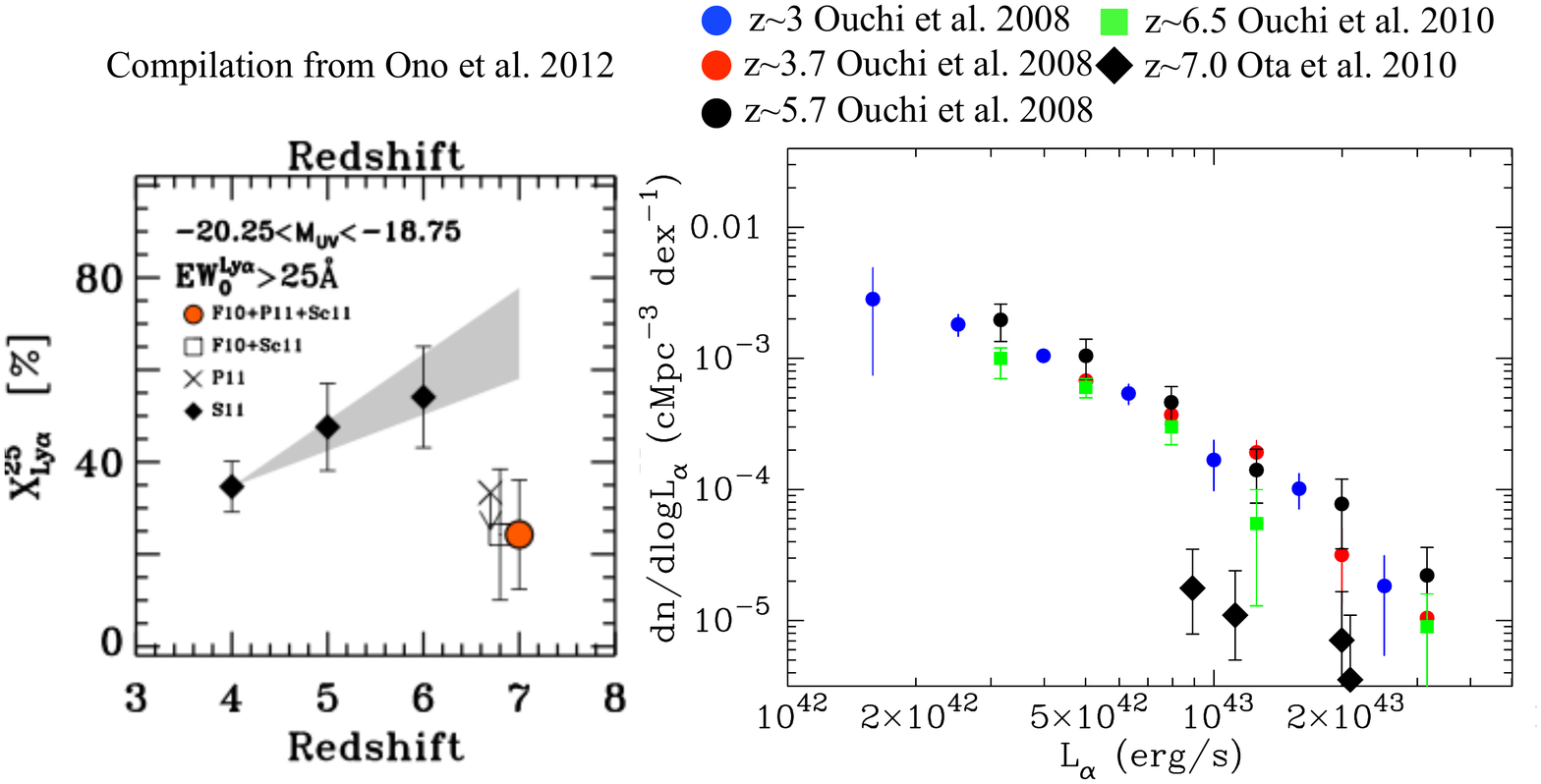}}
\vspace{-20.0mm}
\caption{There are two independent observational indications that the Ly$\alpha$ flux from galaxies at $z>6$ is suppressed compared to extrapolations from lower redshifts. The {\it left panel} shows the drop in the `{\it Ly$\alpha$ fraction}' in the drop-out (Lyman Break) galaxy population, the {\it right panel} shows the sudden evolution in the Ly$\alpha$ luminosity function of Ly$\alpha$ selected galaxies (LAEs). This evolution has been shown to be quantitatively consistent (see Dijkstra \& Wyithe 2012, Gronke et al. 2015).}
\label{fig:obs}       
\end{figure}

{\bf LAE Luminosity Functions.} The Ly$\alpha$ luminosity function (LF) of LAEs barely evolves between $z\sim 3$ and $z\sim 5.7$ \citep{Hu98,Ouchi08}. In contrast, beyond $z\sim 6$ the Ly$\alpha$ luminosity function of LAEs decreases rapidly \citep[][also see the {\it right panel} of Fig~\ref{fig:obs}]{K06,Ouchi10,Ota10,K11,Clement12,Ota12,Jiang13,Faisst14,Konno14}. Importantly, \citet{K06} showed that the (non-ionizing) UV-continuum LF of LAEs did not evolve between $z=5.7$ and $z=6.6$, which suggests that the observed reduction of the Ly$\alpha$ LF is due to a reduction in Ly$\alpha$ flux from LAEs at $z>6$ \cite[also see][]{K11}.\\
{\bf Ly$\alpha$ Fractions.} The `Ly$\alpha$ fraction' of drop-out (LBG) galaxies indicates the fraction of drop-out galaxies for which spectroscopic follow-up reveals a Ly$\alpha$ emission line with EW larger than some threshold value. The Ly$\alpha$ fraction of drop-out galaxies increases from $z\sim 2$ to $z\sim 6$ \citep{Stark10,Stark11,C14}, but then suddenly decreases at $z\sim 7$ \citep[][also see the {\it left panel} of Fig~\ref{fig:obs}]{Fontana10,Pentericci11,Ca12,Ono12,Schenker12,Ca13,Pentericci14} and even more at $z\sim 8$ \citep{Treu12,Treu13}.\\

These two observations are clearly consistent with each other: while UV-LFs of drop-out galaxies decreases continuously with redshift between $z=3-6$, the observation that the Ly$\alpha$ LFs of LAEs do not evolve in the same redshift range, can be explained if the Ly$\alpha$ line becomes increasingly strong between $z=3-6$. \citet{DW12} and \citet{D14} have indeed shown that the observed redshift-evolution of the Ly$\alpha$ fraction in drop-out galaxies and Ly$\alpha$ LFs of LAEs are quantitatively consistent with each other.

\begin{svgraybox} Observations of drop-out galaxies and Ly$\alpha$ emitters independently show that the observed Ly$\alpha$ flux from galaxies at $z>6$ is suppressed significantly when compared to extrapolations from lower redshift data. \vspace{-2mm}\end{svgraybox}

\section{Key Physical Ingredients that Regulate Ly$\alpha$ Visibility}
\label{sec:key}

\begin{figure}[tb]

\includegraphics[scale=.45]{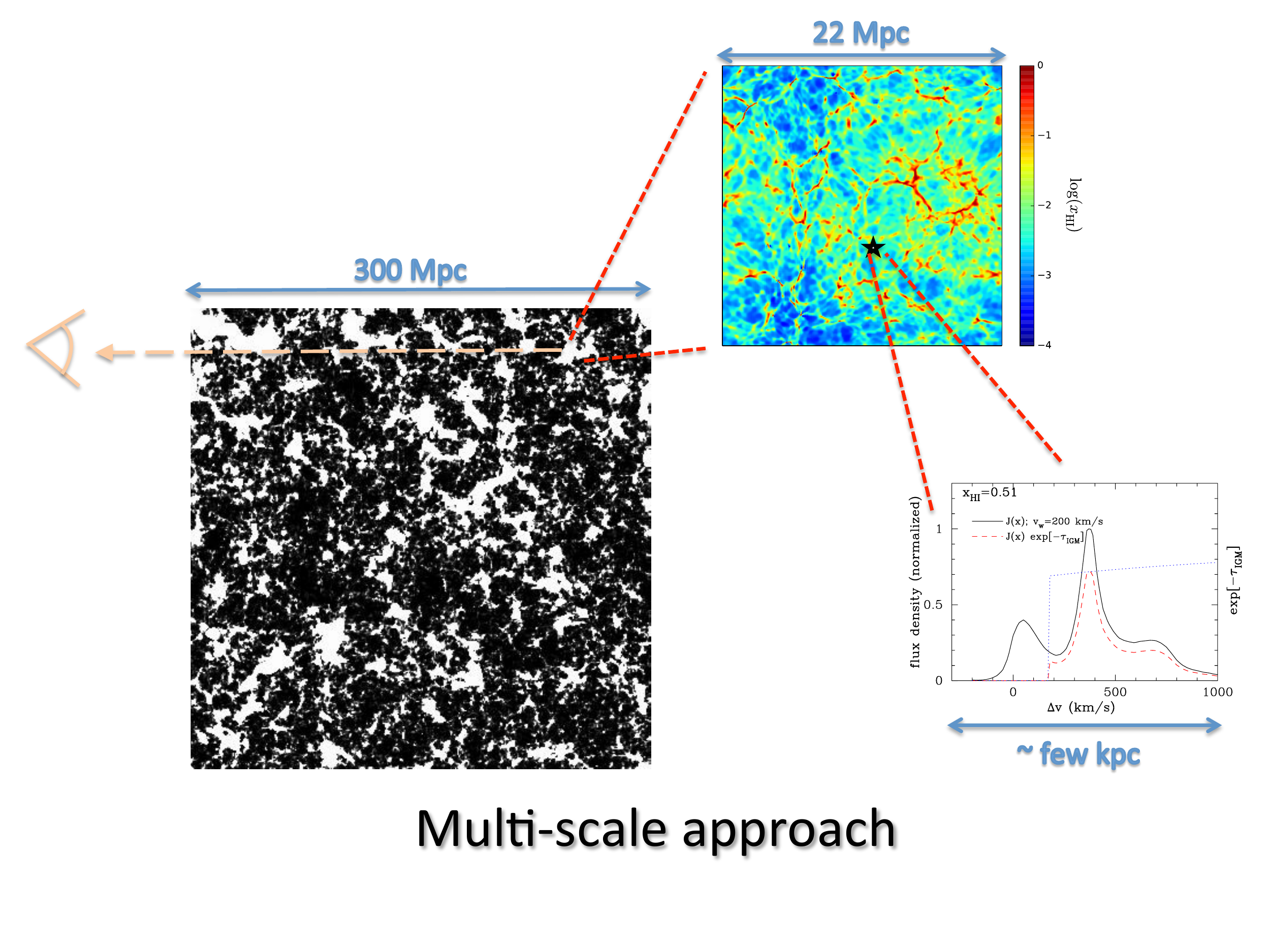}
%
%
\vspace{-11mm}
\caption{Schematic representation of the scales that affect the transfer of Ly$\alpha$ photons during the EoR ({\it Credit: from Figure~1 of Mesinger et al. 2015, `Can the intergalactic medium cause a rapid drop in Ly$\alpha$ emission at $z>6$?', MNRAS, 446, 566}). The {\it left figure} represents that large-scale simulations are required to properly model the diffuse neutral intergalactic component. The {\it upper right figure} shows that ionised bubbles (represented in white in the left figure) are not fully ionised, but instead contain a wealth of structure including self-shielding Lyman-limit systems (indicated in red). On even smaller interstellar scales the Ly$\alpha$ line shape is determined (as represented by the {\it lower right figure}). The line shape can strongly affect the subsequent radiative transfer on larger scales (also see Fig~\ref{fig:igm}).}
\label{fig:1}       
\end{figure}

Ly$\alpha$ transfer occurs on a range of scales: radiative transfer on interstellar scales determines how much Ly$\alpha$ escapes from galaxies, and sets the spectral line profile of escaping Ly$\alpha$ photons. The line profile is important, as it strongly affects the subsequent radiative transfer, and so it provides a key ingredient in the Ly$\alpha$ transfer problem. Here, I discuss key physical processes regulating Ly$\alpha$ transfer in more detail. I have divided this discussion into three parts, each of which corresponds to a range of physical scales. Figure~\ref{fig:1} shows a schematic overview of this discussion. \\

{\bf Transfer in the (Dusty) ISM.} Radiative transfer on interstellar scales is a complex problem as it depends on the kinematic and distribution of neutral gas in the interstellar medium (ISM). A complete review of our current understanding of this process is beyond the scope of this chapter. The interested reader is referred to the reviews by e.g. Dijkstra (2014) and Barnes et al. (2014). Our knowledge can be summarised as follows: ({\it i}) Ly$\alpha$ escape is correlated with dust content: more dust reduces the fraction of Ly$\alpha$ photons that survive their journey through the ISM. The anti-correlation between Ly$\alpha$ escape fraction and dust content has been found in a sample of local galaxies, as well as in Lyman-break galaxies at $z=2-3$. This correlation may be responsible for the observed increase in the Ly$\alpha$ fraction between $z=2-6$ \citep[see e.g.][and references therein]{Hayes11}. The subsequent evolution at $z>6$ cannot be explained with this correlation, as the average dust content of galaxies keeps decreasing \citep[][]{Finkcolors,Bouwenscolor}; ({\it ii}) Ly$\alpha$ escape is facilitated by outflows. Outflows play a major role in determining the Ly$\alpha$ spectral line shape \citep[e.g.][see Dijkstra 2014 for a more complete list of references]{V08}; ({\it iii}) Ly$\alpha$ escape is further facilitated by low column density `holes' in the interstellar medium \citep[e.g.][]{Jones13,Shibuya14}. These three point combined imply that Ly$\alpha$ transfer on interstellar scales is affected strongly by dusty, probably multiphase outflows. This process and its impact on the Ly$\alpha$ spectral line profile is represented by the {\it lower right panel} in Figure~\ref{fig:1}.\\

{\bf Transfer in the ionised IGM/CGM.} Studies of quasar absorption lines (i.e. the `Ly$\alpha$ forest') indicate that the ionised IGM at $z\gsim 4$ is optically thick to all photons emitted short-ward of the Ly$\alpha$ resonance\footnote{The `effective' optical depth in the Ly$\alpha$ forest exceeds unity, $\tau_{\rm eff} > 1$ at $z \gsim 4$ \citep[see e.g Fig~3 of][]{FG08}}. This implies that to first order, the ionised IGM transmits all flux redward of the Ly$\alpha$ resonance, and suppresses the flux on the blue side of the resonance. In this first-order estimate, the IGM transmits $\mathcal{T}_{\rm IGM} > 50\%$ of all Ly$\alpha$ photons.
\begin{figure}[tb]
\includegraphics[scale=.40]{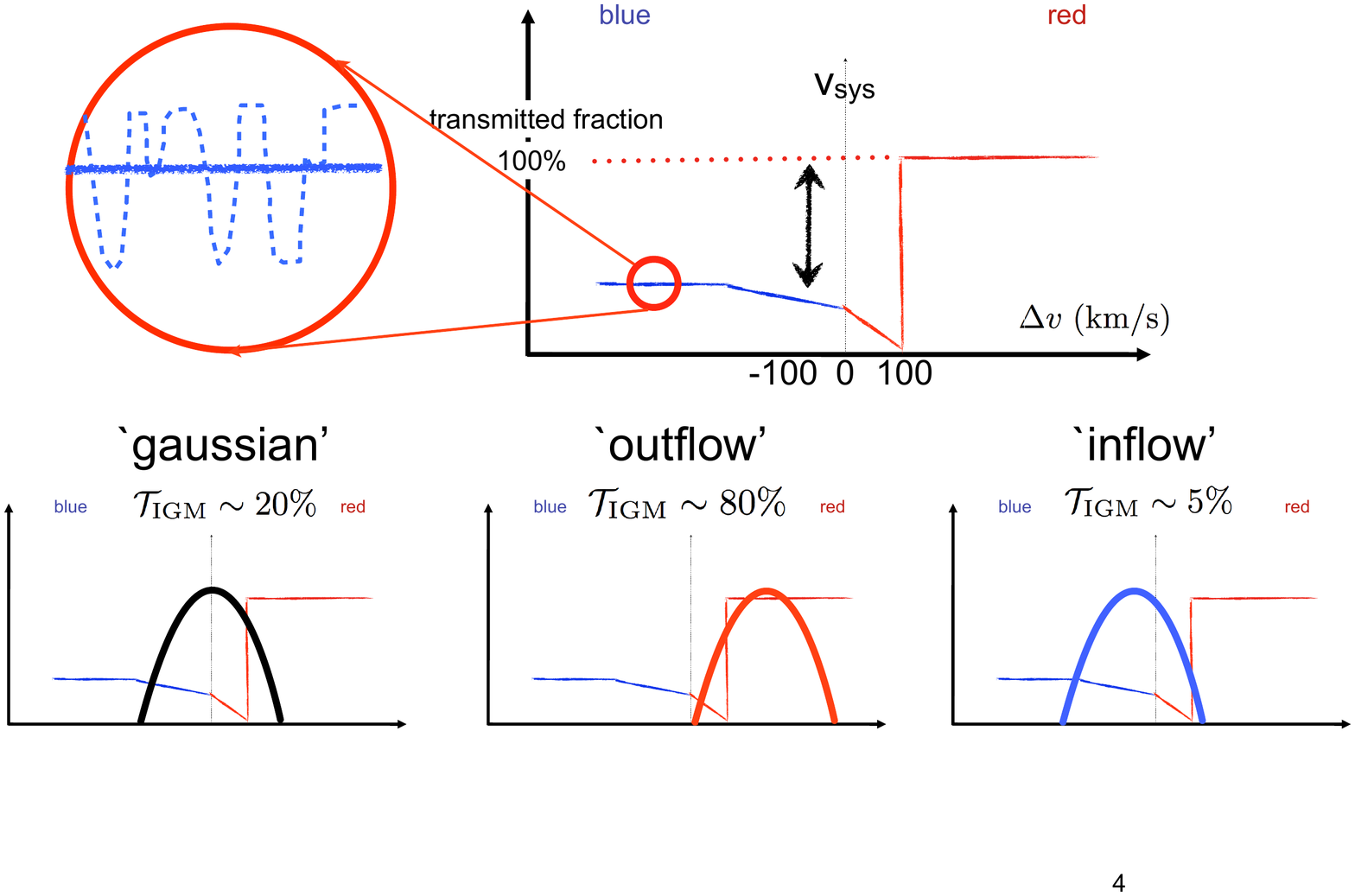}
\vspace{-15mm}
\caption{Schematic representation of the impact of a residual amount of intergalactic HI on the fraction of photons that is directly transmitted to the observer, $\mathcal{T}_{\rm IGM}$. The {\it top panel} shows that the Ly$\alpha$ forest (shown in the {\it inset}) suppress flux on the blue side of the Ly$\alpha$ line. Overdense gas in close proximity to the galaxy - this gas can be referred to as `circum-galactic gas - enhances the opacity in the forest at velocities close to systemic (i.e $v_{\rm sys}=0$). Inflowing circum-galactic gas gives rise to a large IGM opacity even at a range of velocities redward of the Ly$\alpha$ resonance. The {\it lower panels} show that the total faction of photons transmitted though the IGM, $\mathcal{T}_{\rm IGM}$, depends strongly on the assumed line profile, which is represented by the {\it thick solid lines}. The {\it lower left figure} shows that for lines centered on $v_{\rm sys}=0$ (here symmetric around $v_{\rm sys}=0$ for simplicity), the IGM cuts off a significant fraction of the blue half of the line, and some fraction of the red half of the line. For lines that are redshifted [blueshifted] w.r.t $v_{\rm sys}$, larger [smaller] fraction of emitted Ly$\alpha$ photons falls outside of the range of velocities affected by the IGM. The line profiles thus plays a key role in the intergalactic radiative transfer process.}
\label{fig:igm}       
\end{figure}

However, radiative transfer in the IGM is more complex than suggested by this first-order estimate: intergalactic gas in close proximity to the galaxy (this gas is also referred to as the `circumgalactic medium') is significantly more dense than average \citep[see e.g. Fig~2 of][]{Barkana04}. Moreover, the gravitational potential of the dark matter halo hosting Ly$\alpha$ emitting galaxies can alter the velocity field of this circumgalactic gas \citep[e.g.][]{Barkana04}. Both effects combined cause the residual neutral hydrogen in the ionised gas to be opaque to Ly$\alpha$ photons, especially at frequencies corresponding to small velocity off-sets from the systemic velocity of the galaxy \citep[][]{Santos04,IGM,Laursen11}. This `thickening' of the Ly$\alpha$ forest around the Ly$\alpha$ line is illustrated {\it top panel} of Figure~\ref{fig:igm}, which (schematically) shows the fraction of photons that can propagate to the observer without being scattered, as a function of velocity off-set ($\Delta v$) from a galaxies systemic velocity. This plot illustrated three points: ({\it i}) at large velocity off-sets on the blue-side of the line, the IGM transmits a constant (small) fraction of the Ly$\alpha$ flux. This fraction has been measured in quasar absorption line studies, ({\it ii}) as we approach $\Delta v=0$ from the blue side, the Ly$\alpha$ forest thickens (as mentions above), ({\it iii}) this thickening can extend up to $\sim 200$ km s$^{-1}$ to the red side of the Ly$\alpha$ resonance (where the precise number is close to the circular velocity of the dark matter halo that hosts the galaxy), which is due to in falling overdense gas \citep[][]{Santos04,IGM,Laursen11}.  Along these lines \citet{Finkcolors} propose that the observed reduction in Ly$\alpha$ flux may be related to the ratio of the gas accretion rate onto galaxies and their star formation rate, which has been inferred observationally to increase by $\sim 40\%$ from $z=6$ to $z=7$ \citep{Papovich11}. 

It has been shown that as a consequence of these more complicated processes, the IGM can transmit as little as $\mathcal{T}_{\rm IGM} \sim 10-30\%$ of all Ly$\alpha$ photons, even when it was highly ionised \citep{IGM,Iliev08,Dayal11,Laursen11}. As the mean density of the Universe increases as $(1+z)^3$, we expect the CGM/IGM to become increasingly dense \& potentially more opaque towards higher redshift \citep{LF,Laursen11}. Observations of Ly$\alpha$ halos around star forming galaxies provide hints that scattering in this CGM may be more prevalent at $z=6.5$ than at $z=5.7$, although the statistical significance of this claim is weak \citep{Momose14}.

Finally, it is important to realise that the transmitted fraction of Ly$\alpha$ photons, $\mathcal{T}_{\rm IGM}$, depends strongly on the assumed line-shape \citep[also see][]{Haiman02,Santos04}. The fraction $\mathcal{T}_{\rm IGM}$ is the integral over the velocity-dependent transmission (shown in Fig~\ref{fig:igm}) weighted by the line flux density as a function of velocity. If the emerging Ly$\alpha$ line is centered on the systemic velocity, then $\mathcal{T}_{\rm IGM}=10-30\%$. However, if the line is blue shifted (as a result of scattering through optically thick collapsing gas) or redshifted (as a result of scattering through optically thick outflowing gas), then the total transmitted fraction may be affected significantly. This is illustrated by the {\it lower three panels} in Figure~\ref{fig:igm}. {\it That is, not only do we care about how much Ly$\alpha$ escapes from the dusty ISM, we must care as much about how the emerging photons escape in terms of the line profile}.\\

\begin{figure}[tb]
\vspace{-9mm}
\includegraphics[scale=.38]{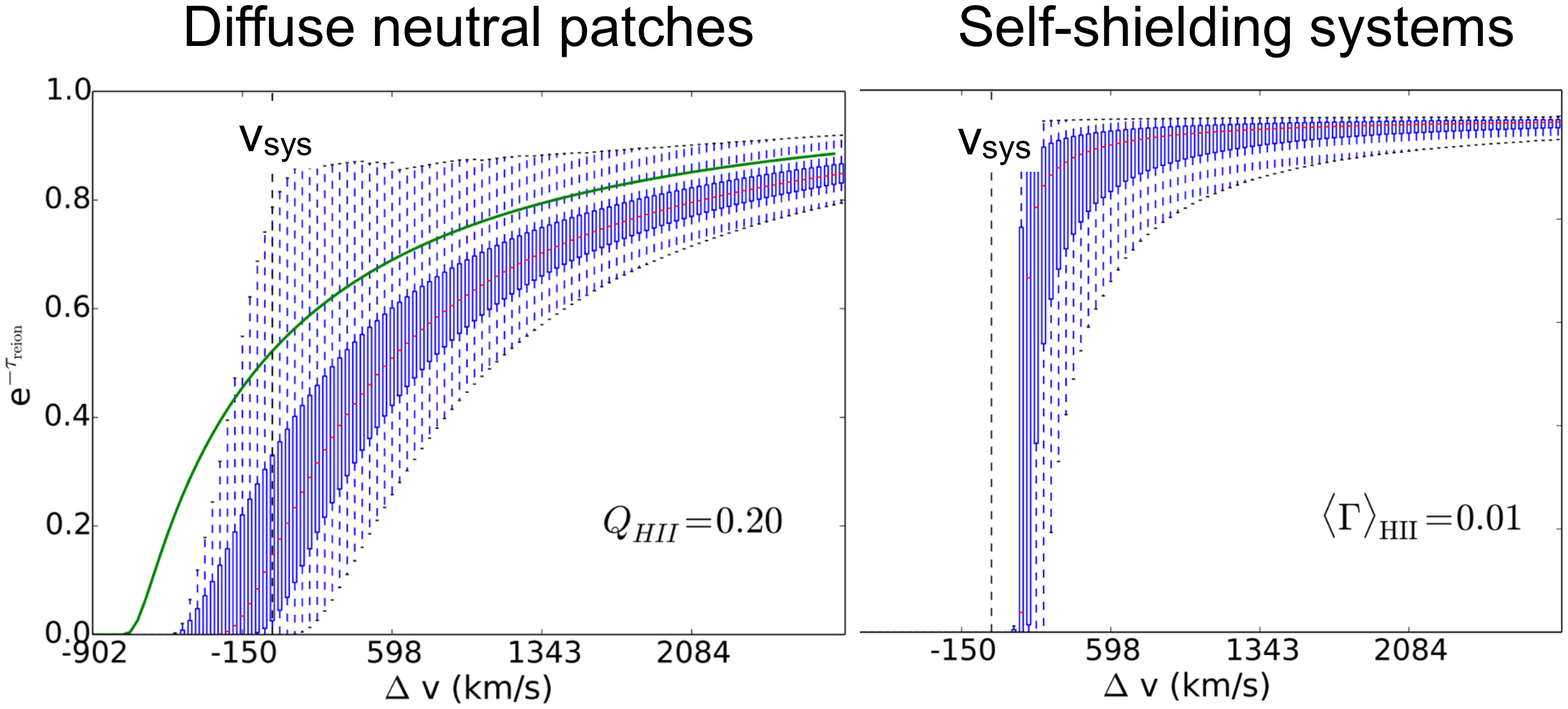}
\vspace{-30mm}
\caption{Neutral gas in the intergalactic medium can give rise to a large `damping' wing opacity that extends far to the red side of the Ly$\alpha$ resonance. The {\it left panel} [{\it right panel}] shows the IGM transmission (as in Fig~\ref{fig:igm}) as a function of velocity off-set for the diffuse neutral IGM (for $x_{\rm HI}=0.8$)[self-shielding gas clouds] ({\it Credit: from Figures~2 and ~4 of Mesinger et al. 2015, `Can the intergalactic medium cause a rapid drop in Ly$\alpha$ emission at $z>6$?', MNRAS, 446, 566}). To obtain the {\it total} IGM transmission, one should multiply the transmission curves shown here and in Fig~\ref{fig:igm}.}
\label{fig:igmdamp}       
\end{figure}
In addition to this residual HI in the ionised IGM, the ionised IGM contains overdense gas which can self-shield and form Lyman limit systems (LLSs, $N_{\rm HI} > 10^{17}$ cm$^{-2}$). The optical depth through these systems to Ly$\alpha$ photons can be significant even far from line centre, $\tau \sim 1(N_{\rm HI}/4 \times 10^{18}\hs{\rm cm}^{-2})(\Delta v/100\hs{\rm km/s})^{-2}$, which makes them opaque even in the damping wing of the Ly$\alpha$ absorption cross-section. This off-resonance opacity is often referred to as the `damping wing opacity', and is shown in Figure~\ref{fig:igmdamp}. The damping wing opacity has a different dependence on $\Delta v$ compared to that of the ionised IGM (which was shown in Fig~\ref{fig:igm}): most notably the power-law dependence of the Ly$\alpha$ absorption cross-section in its damping wings cause the transmission to increase continuously with $\Delta v$, instead of the abrupt increase in $\mathcal{T}({\Delta v})=1$ due to the ionised IGM redward of some critical velocity. 

These self-shielding systems can provide a significant source of opacity during the EoR, and may even fully explain the observed drop in the Ly$\alpha$ fraction at $z>6$ \citep{BH13}, though this requires a drop in the ionising background by a factor of $\sim 30$ \citep[see the discussion in][]{Mesinger14}. Self-shielding gas is represented as the red patches in the {\it upper right panel} of Figure~\ref{fig:1}. Note that the thickening of the forest at small velocity off-sets from systemic occurs on smaller scales closer to the galaxy (i.e. circum-galactic scales).\\

{\bf Transfer in the diffuse neutral IGM.} During the reionization process spatial fluctuations existed in the neutral fraction of hydrogen over scales exceeding tens of cMpc. These large-scale fluctuations are a consequence of the very strong clustering of the first generations of galaxies. In order to capture these large spatial variations, one needs simulations exceeding 100 cMpc in size \citep[see e.g.][for a review]{Trac11}. The {\it left panel} in Figure~\ref{fig:1} shows a snap-shot from a simulation that was 300 cMpc on-a-side. White [black] patches indicate fully ionised [neutral] hydrogen. Figure~\ref{fig:1} clearly shows the inhomogeneous, or ÒpatchyÓ, nature of the reionization process. Ionized bubbles first emerge in overdense regions of the Universe. Ly$\alpha$ emitting galaxies that are luminous enough to be detectable with existing instruments likely populated the more massive dark matter halos that existed at these epochs, with $M_{\rm DM} \gsim 10^{10} M_{\odot}$. These dark matter halos preferentially resided in overdense regions, and hence inside the ionised bubbles. Ly$\alpha$ emitting galaxies thus likely resided predominantly inside ionised bubbles. Ly$\alpha$ photons emerging from these galaxies can therefore redshift far away from resonance before encountering the neutral IGM during their flight their the ionised bubbles\footnote{Photons that do scatter in the diffuse neutral IGM form so called `Loeb-Rybicki halos' \citep{LR99}, which are several orders of magnitude fainter than the currently observed halos \citep[see Fig~A1 of][]{DW10}.}. Patchy (or inhomogeneous) reionization therefore weakens the impact of the diffuse neutral IGM on the observed Ly$\alpha$ flux from galaxies \citep[especially at the later stages of reionization, ][]{Madau00,Haiman02,Gnedin04,Furlanetto04,Furlanetto06,McQuinn07,Mesinger08,Iliev08,Dayal11,D11,Jensen,Hutter14,Choudhury15}.

\vspace{-7mm}
\section{Interpreting the Observations}
\label{sec:interpret}
\vspace{-2mm}
Despite the fact that Ly$\alpha$ transfer is a multi-scale problem which is not well understood (as discussed above), reionization should be considered the prime candidate to explain the observed reduction in Ly$\alpha$ flux from galaxies at $z>6$: radiative transfer processes on interstellar, circumgalactic \& intergalactic scales together increase the visibility of Ly$\alpha$ flux from galaxies from $z=0$ to $z=6$ \citep[see e.g][]{Hayes11}. A simple physical picture can explain this observation: dusty outflows regulate the escape of Ly$\alpha$ photons from galaxies. As the average dust content of galaxies decreases with redshift, the escape fraction increases. Moreover, the presence of outflows causes the Ly$\alpha$ line to emerge with a redshift, which makes the photons less susceptible to scattering in the CGM/ionized IGM\footnote{This reduced sensitivity to CGM/IGM opacity is important as we expect its opacity to Ly$\alpha$ photons to {\it increase} with redshift, while observations indicate it is increasingly easy to detect Ly$\alpha$ flux from galaxies towards higher redshifts from $z=0$ to $z=6$.} (this situation is illustrated in the {\it central lower panel} of Fig~\ref{fig:igm}). 

At $z>6$ suddenly the observed trend reverses, exactly when other observations indicate that we may be entering the epoch of reionization. 
It is therefore very reasonable to ask: `what if the observed reduction in Ly$\alpha$ flux from galaxies at $z>6$ is due to reionization, then what does this say about reionization?' \\

This question is easiest to address with the observed evolution of the Ly$\alpha$ fraction. Here, we need to assign a functional form of the EW-PDF at $z=6$ that is consistent with observations, and then ask how reionization affects this PDF at $z>6$. The EW-PDFs are related via the $\mathcal{T}_{\rm IGM}$-PDF as \citep{D11}
\begin{equation}
P_7({\rm EW}) \propto \int d\mathcal{T}_{\rm IGM}\hs P_7(\mathcal{T}_{\rm IGM})P_6({\rm EW}/\mathcal{T}_{\rm IGM}),
\end{equation} where $P_7$ [$P_6$] denotes the Ly$\alpha$ EW PDF at $z=7$ [$z=6$], and $P_7(\mathcal{T}_{\rm IGM})$ denotes the fraction of Ly$\alpha$ photons that is transmitted through the neutral IGM at $z=7$. It is useful to explicit provide the expression for $\mathcal{T}_{\rm IGM}$, as it encodes the three ranges of scales indicated in Figure~\ref{fig:1} \citep{Mesinger14}:

\begin{eqnarray}
\label{eq:tauigm}
\mathcal{T}_{\rm IGM} = \int_{-\infty}^{\infty}d\Delta v\hs J_{\alpha}(\Delta v) \exp [-\tau_{\rm IGM}(z_{\rm g},\Delta v)], \\ \nonumber
\tau_{\rm IGM}(z_{\rm g},\Delta v)=\taud(z_{\rm g},\Delta v)+\tauHII(z_{\rm g},\Delta v).
\end{eqnarray} Here, $J_{\alpha}(\Delta v)$ denotes the line profile of Ly$\alpha$ photons as they escape from the galaxy, and thus encodes radiative transfer on the smallest (i.e. interstellar) scales. Eq~\ref{eq:tauigm} shows explicitly that there are two components to the IGM opacity: ({\it i}) $\taud(z_{\rm g},\Delta v)$ describes the opacity in diffuse neutral intergalactic patches on the largest scales, and this component is thus unique to the reionization epoch, and ({\it ii}) $\tauHII(z_{\rm g},\Delta v)$ the opacity in the ionized component of the IGM/CGM (this includes LLSs that reside inside the ionised bubbles). Most analyses have ignored this last component, which is reasonable for scenarios in which (dusty) outflows regulate Ly$\alpha$ escape from galaxies (as is likely the case at $z=0-6$, see \S~\ref{sec:key}).

Early results following this line of reasoning conclude that {\it if } the entire drop in the Ly$\alpha$ fraction is due to diffuse neutral intergalactic values, then the data at face value implies that the volume-averaged neutral fraction\footnote{The quantity $x_{\rm HI}$ will refer to the volume averaged neutral fraction of hydrogen throughout this chapter.} is $x_{\rm HI}(z=7)\gsim 0.5$ \citep{D11,Jensen}. 
This constraint in consistent with earlier constraints on reionization from the observed evolution of the Ly$\alpha$ luminosity function of LAEs \citep{McQuinn07}. This agreement is a consequence of the fact that the observed drop in the Ly$\alpha$ fraction at $z>6$ is quantitatively consistent with the observed evolution in the LAE Ly$\alpha$ luminosity function \citep{D14}.\\

\begin{figure}[tb]
\vspace{-5mm}
\includegraphics[scale=.33]{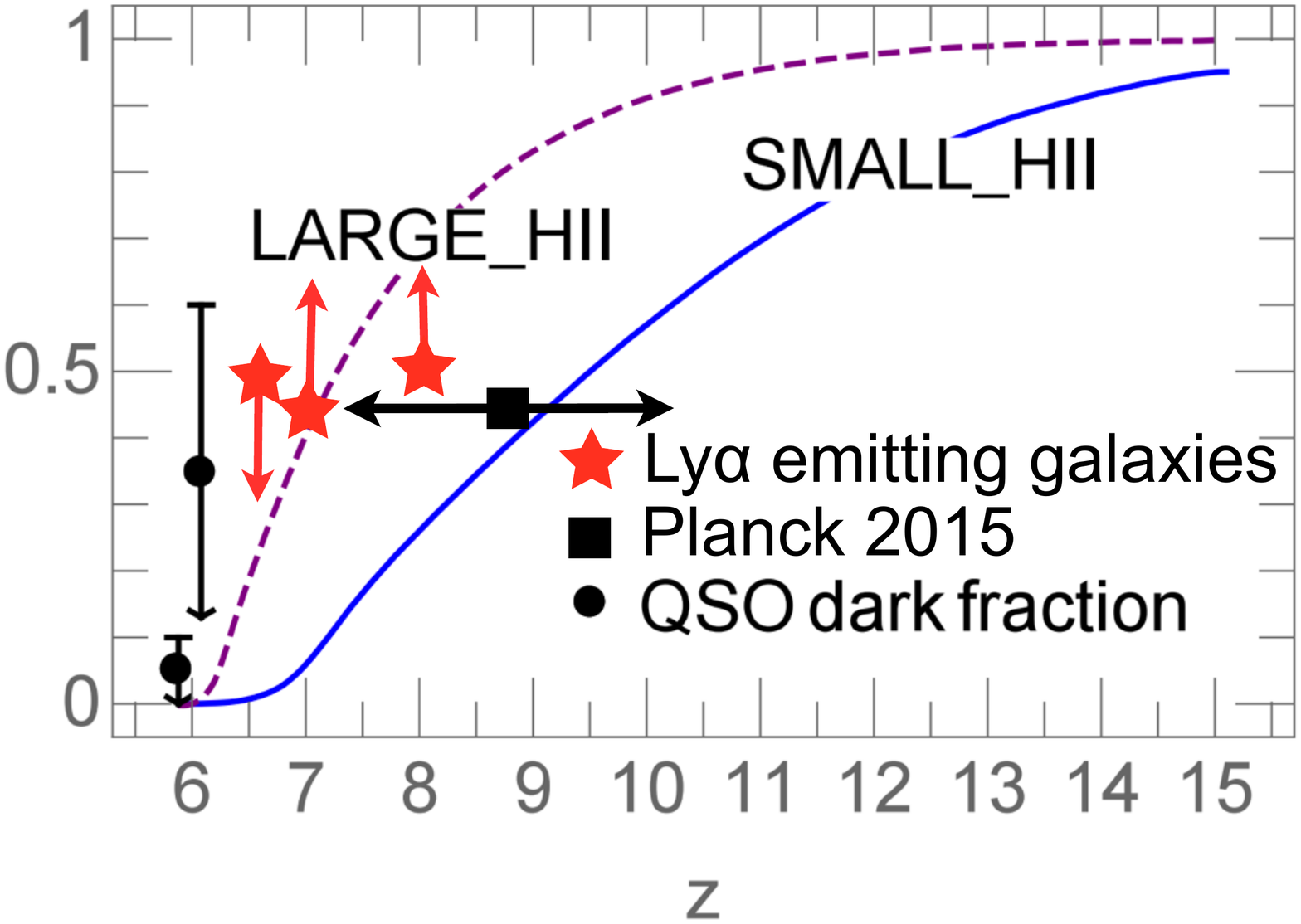}
\vspace{-5mm}
\caption{Current observational constraints on the volume filling factor of the neutral IGM as a function of redshift. The {\it filled black circles} represent mode lindependent upper limits around $z \sim 6$ from the dark fraction in quasar absorption spectra (McGreer et al. 2015). The {\it filled black square} shows the constraint obtained by the Plank satellite. The {\it red stars} indicate constraints from Ly$\alpha$ emitting galaxies. The lower limits at $z=7$ and $z=8$ are derived from the observed drop in `Ly$\alpha$ fraction' of drop-out galaxies, while the lower limit at $z=6.6$ is derived from the observed clustering of LAEs ({\it Credit: Figure is adapted from Figure 3 of Sobacchi \& Mesinger 2015, ÔThe clustering of  Ly$\alpha$ emitters at $z \approx7$: Implications for Reionization and Host Halo MassesÕ, arXiv:1505.02787}). Current observational constraints are all consistent the reionization process lasting down to $z\sim 6$.}
\label{fig:xhi}       
\end{figure}

The required change in $x_{\rm HI}$ can in theory be reduced significantly if one accounts for the redshift evolution of the opacity of the IGM in ionised bubbles (i.e. $\tauHII$ in Eq~\ref{eq:tauigm}). \citet{BH13} have shown that self-shielding gas in LLSs and DLAs can fully account for the observed drop in the Ly$\alpha$ fraction, and that the volume filling factor of their neutral gas would only be $x_{\rm HI}\sim 0.05-0.1$ at $z=7$. However, Mesinger et al. (2015) have recently shown that in order for LLSs to fully explain the observed drop in the Ly$\alpha$ fraction, the ionising background inside HII bubbles must fall by a factor of $\sim 30$, which they consider unreasonable. Mesinger et al. (2015) present constraints on $x_{\rm HI}$ marginalised over the ionising background, and conclude that\footnote{During the final stages of preparation, a preprint by \citet{Choudhury15} appeared which constrained $x_{\rm HI}\sim 0.3$ at $z\sim 7$ for a model that is similar to that of Mesinger et al. (2015).  \citet{Choudhury15} adopt a steeper EW-PDF $P_6({\rm EW})$, which makes all Ly$\alpha$ emitting galaxies fainter by a factor of $\sim 0.8$, and could explain their somewhat smaller required $x_{\rm HI}$. This further illustrates how current observational uncertainties on Ly$\alpha$ EW-PDFs at $z\sim6$ and $z\sim7$ limit our ability to constrain $x_{\rm HI}$.} $x_{\rm HI}(z=7)>0.4$. This constraint is shown in Figure~\ref{fig:xhi}, where it is compared to constraints from other probes of reionization. \citet{D14} recently showed that this constraint reduces to $x_{\rm HI}(z=7)>0.2$ if one extrapolates the inferred redshift evolution of ionising photons ($f_{\rm esc}^{\rm ion}$, see e.g. Inoue et al. 2006, Kuhlen \& Faucher-Giguere 2012, Becker \& Bolton 2013) at $z \leq 6$ to $z=7$. Since the amount of {\it produced} Ly$\alpha$ emission scales as $1-f_{\rm esc}^{\rm ion}$, this can help explain the observed reduction in Ly$\alpha$ flux. It is important to stress that the quoted limits on $x_{\rm HI}(z=7)$ are based on the median of the observed data at $z=7$. The limited size of the sample of drop-out galaxies with spectra at $z \geq 6$ translates to large uncertainties in $x_{\rm HI}(z=7)$ \citep[e.g.][]{Taylor13,Mesinger14}: \citet{Mesinger14} notes that formally $x_{\rm HI}=0.0$ is consistent with the Ly$\alpha$ fraction data at 95\% CL (though $x_{\rm HI}=0.0$ would be ruled out with greater significance if additional constraints from the observed evolution of the Ly$\alpha$ luminosity functions of LAEs were included).\\

This section can be summarised as follows:
\begin{svgraybox}
While Ly$\alpha$ radiative transfer is a complex, multi scale process, current observations strongly suggest that Ly$\alpha$ flux from galaxies at $z>6$ is suppressed by additional intervening neutral hydrogen atoms, with current models favouring a substantial neutral fraction at $z\sim 7$. It remains to be seen how much this constraint is affected by ({\it i}) residual HI inside the `circum-galactic medium', or in self-shielding Lyman-limit systems populating the ionised IGM, and ({\it ii}) the possible redshift evolution of galaxy properties - most notably the ionising photon escape fraction.
\end{svgraybox} 
In the next section, I discuss how we may be able to tell apart these scenarios observationally.
\vspace{-6mm}
\section{How Future Surveys will probe Reionization}
\label{sec:future}
\vspace{-3mm}
\subsection{$z$-Evolution of Ly$\alpha$ Luminosity Functions and Ly$\alpha$ Fractions}

Future instruments and surveys will provide us with much better constraints on the redshift evolution of the Ly$\alpha$ fraction and Ly$\alpha$ luminosity functions. This is important: reducing for example the uncertainties on the observed Ly$\alpha$ fractions at $z=6$ and $z=7$ by a factor of $\sim 2$ would already allow for much more stringent joint constraints \citep{Mesinger14}.  There are several new instruments that will contribute here:

\begin{itemize}
\item MUSE\footnote{http://www.eso.org/sci/facilities/develop/instruments/muse.html} on VLT  is sensitive to fluxes of $F_{{\rm Ly}\alpha}\sim 4 \times 10^{-19}$ erg s$^{-1}$ cm$^{-2}$ (in Wide Field Mode, after 80hrs of observation). This flux corresponds to a luminosity $L_{\alpha}\sim 1.6\times 10^{41}$ erg s$^{-1}$ at $z=6$. We therefore expect MUSE to be better constrain the Ly$\alpha$ EW-PDF (and therefore the Ly$\alpha$ fractions) over the redshift range $z=3-7$. Similarly, current Ly$\alpha$ luminosity functions of LAEs are well constrained only over a range of luminosity of $\log L_{\alpha}\sim 42-43$. MUSE is capable of significantly extending this range.

\item The `Keck Cosmic Web Imager'\footnote{http://www.srl.caltech.edu/sal/keckcosmic-web-imager.html} which is designed to perform high-precision spectroscopy on faint objects, which include Ly$\alpha$ emitting galaxies at $5 \lsim z \lsim 7$ \citep[e.g][]{Martin10}.

\item Spectroscopic observations of intrinsically faint, gravitationally lensed galaxies provide complementary constraints on faint Ly$\alpha$ emitting galaxies \citep[e.g.][]{Stark07}. For example, the currently ongoing Grism Lens-Amplified Survey from Space (GLASS) will obtain spectra of galaxies in the core and infall regions of 10 galaxy clusters to look for line emission from gravitationally lensed high-redshift galaxies \citep{Schmidt14}. 

\item Finally, the new Hyper Suprime-Cam\footnote{http://www.naoj.org/Projects/HSC/} on the Subaru telescope has a field-of-view with a diameter of 1.5$^{\circ}$. With this camera, planned surveys on Subaru covering tens of deg$^2$ will increase the sample of LAEs at $z=6.5$, $z=7.0$, and $z=7.3$ by one (possibly two) orders of magnitude. Increased sample size of LAEs will allow us to investigate the impact of reionization on the clustering properties of LAEs (see \S~\ref{sec:clustering})

\end{itemize}

The redshift evolution of the Ly$\alpha$ Luminosity Functions and Ly$\alpha$ Fractions will provide constraints on $x_{\rm HII}$, but these constraints will be degenerate at some level with the average photoionisation rate inside the HII regions $\Gamma$ \citep[see e.g. Fig ~7-9 in][]{Mesinger14}. This degeneracy can likely be broken with clustering measurements which are discussed next.

\subsection{LAE Spatial Clustering}
\label{sec:clustering}

\begin{figure}[tb]
\vspace{-20.0mm}
\centerline{\includegraphics[scale=.27]{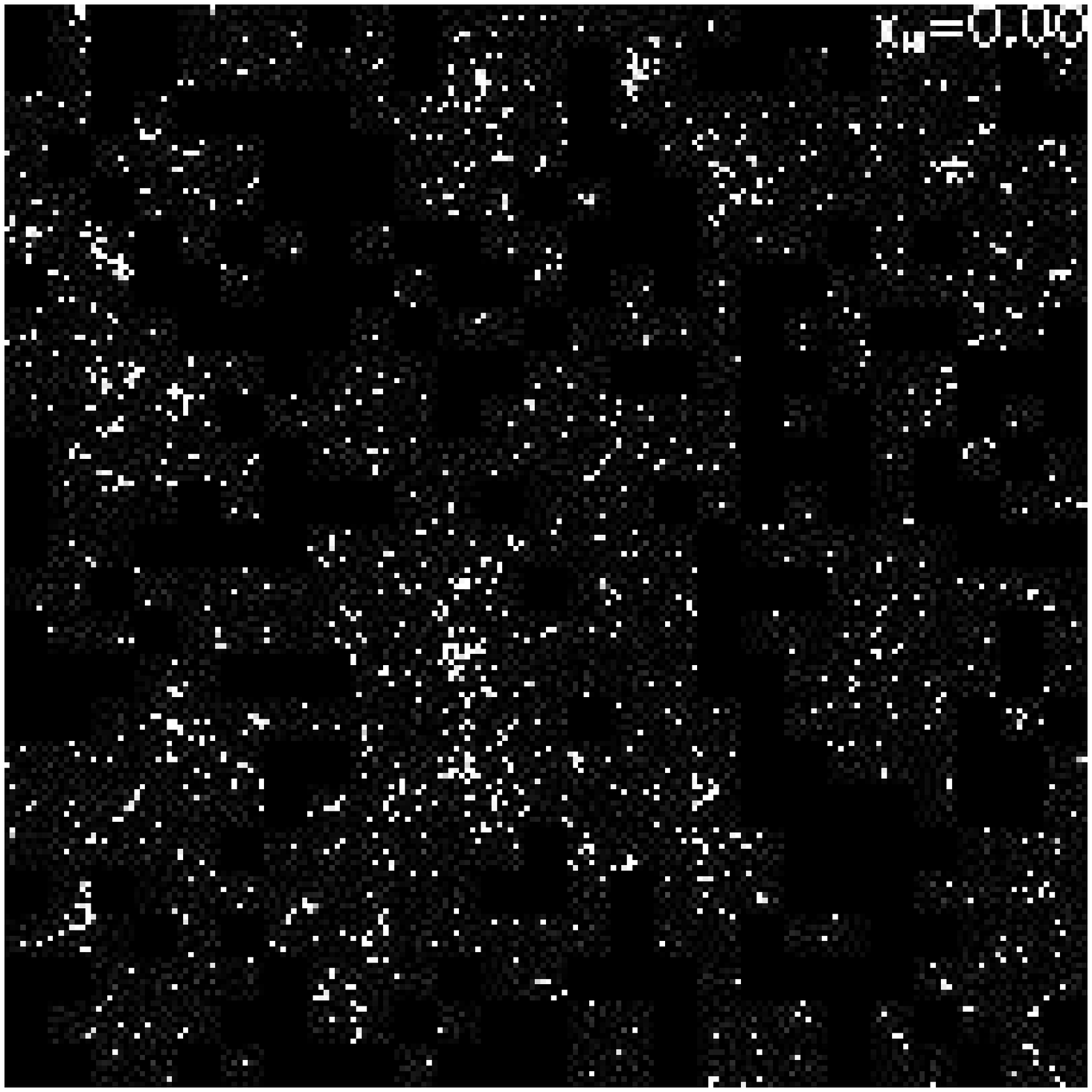}\includegraphics[scale=.27]{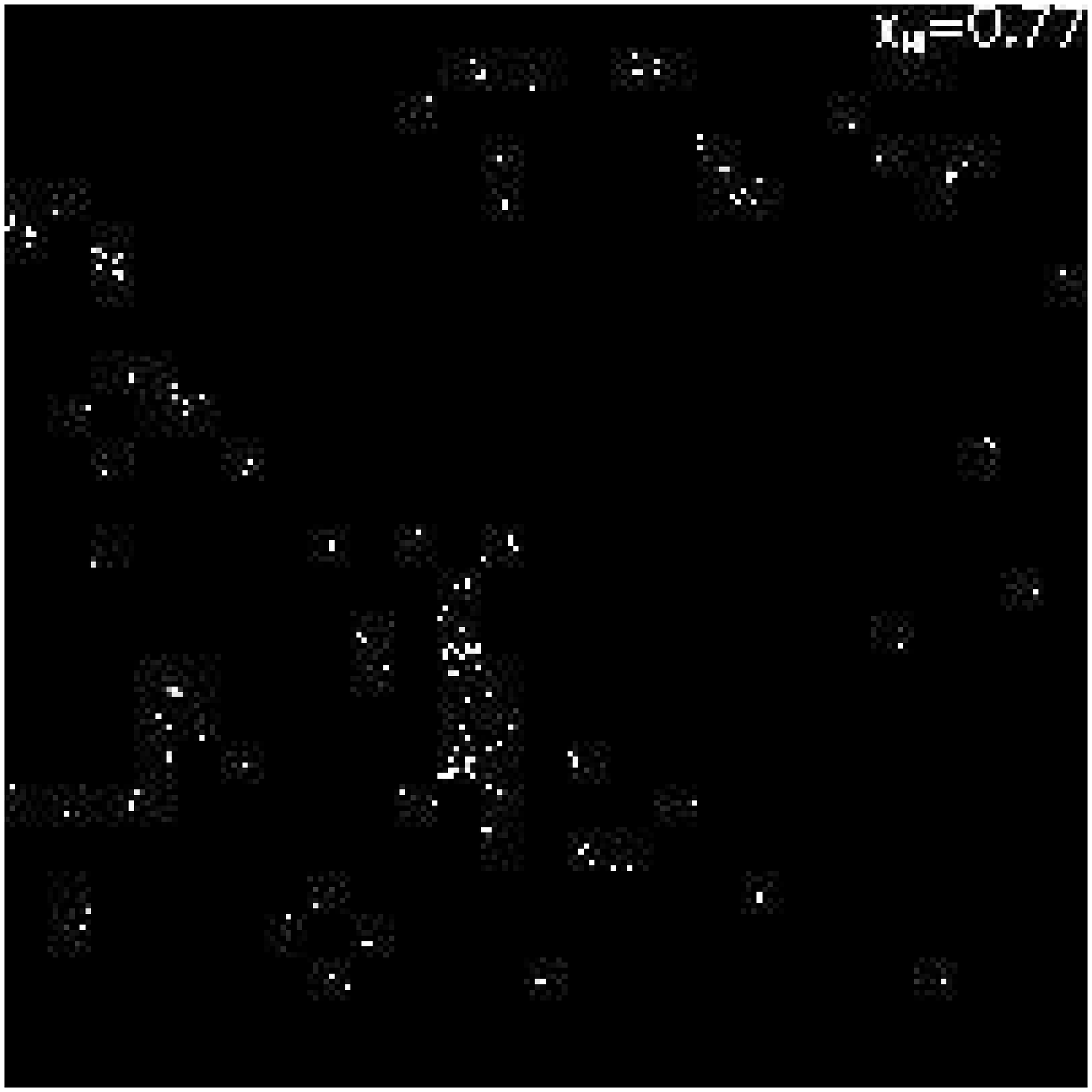}}
\vspace{0.0mm}
\caption{This Figure shows how neutral intergalactic gas modulates the observed {\it projected} distribution of LAEs on the sky. Both panels are $250$ cMpc on each side, and were constructed from a slice that was 20 cMpc deep (which corresponds to a typical width of a narrowband filter of $\sim 100$ \AA). The {\it left panel}/[{\it right panel}] shows the predicted distribution of LAEs without a neutral IGM/with a an IGM for which $x_{\rm HI}=0.77$ ({\it Credit: from Figure~1 of Mesinger \& Furlanetto 2008b, `Ly$\alpha$ Emitters During the Early Stages of Reionization', MNRAS, 386, 1990M}). }
\label{fig:igm2}       
\end{figure}

During reionization the neutral fraction of hydrogen fluctuates over scales exceeding tens of cMpc. We therefore expect (order unity) fluctuations in the observed Ly$\alpha$ flux from galaxies over similar scales (see Fig~\ref{fig:igm2}). The impact of reionization on observed clustering of LAEs has been quantified in two ways:\\

{\bf Angular Clustering.} The angular two-point correlation function $w(\theta)$ denotes the excess probability over random of finding two Ly$\alpha$ emitting galaxies at angular separation $\theta$. \citet{K06} presented measurements of the clustering of LAEs at $z=6.6$. \citet{McQuinn07} compared these measurements to predictions they obtained from a 200 cMpc radiative transfer simulation of cosmic reionization, which they populated with LAEs. \citet{McQuinn07} concluded that the observed clustering measurements imply that $x_{\rm HI} < 0.5$ (2$\sigma$, also see Sobacchi \& Mesinger 2015). \citet{Jensen} recently showed that this constraint relaxes somewhat if one accounts for a scatter in halo mass in intrinsic Ly$\alpha$ luminosity. \citet{Jensen} quantified the significance with which one can detect a reionization signature on LAE clustering as a function of both $x_{\rm HI}$ and the number of LAEs in the survey. For example, for $N_{\rm LAE}=10^3$ (an order of magnitude increase) should be able to detect a reionization signature at $>2 \sigma$ if $x_{\rm HI} > 0.5$. The significance appears to double after $N_{\rm LAE}=500$ which suggests that if Hyper Suprime-Cam would detect a reionization signature robustly if it were to detect $\gg 10^3$ LAEs at $z\sim 7$ (the precise significance depends on the mass range of the halos hosting LAEs, see Fig~7 of Sobacchi \& Mesinger 2015). \\
{\bf Count-in-cells.} \citet{MF08} studied the different `count-in-cell' statistic, which represented integrated measurements of the full clustering (also see Jensen et al. 2014). During reionization we expect the excess probability of finding another LAE in the field that contains an LAE to be higher than post-reionization. This count-in-cell statistic has the advantages that it is easier to generate from a small sample of LAEs, and that it is easier to interpret in non-uniform surveys and/or in small deep fields such as those obtained with e.g. the James Webb Space Telescope. \citet{MF08} predict that in a Universe with $x_{\rm HI}>0.5$ the likelihood of finding more than one LAE among a subset of fields which contain LAEs is $\gsim 10\%$ greater than in the Universe with $x_{\rm HI}=0.0$, and show that this effect could be detected with a few tens of 3-arcminute cubical cells. \\

The count-in-cell and angular clustering measurements require different survey strategies, and clearly complement each other well.We expect processes other than reionization that reduce Ly$\alpha$ flux from galaxies at $z>6$ (see \S~\ref{sec:key}) to leave a signature on LAE clustering that differs from the reionization signature. Quantitative predictions on how these other processes\footnote{The CGM may be more opaque in overdense regions of the Universe, which may make it more difficult to see LAEs in overdense regions. Thus the opacity of the CGM could counteract galaxy bias. This effect has been studied in detail post-reionization by e.g \citet{Zheng11,WD11,Behrens13}. } affect the observed clustering of LAEs at $z>6$ do not exist yet.
\vspace{-5mm}
\subsection{LAE-21cm Cross-Correlation}
Directly detecting the redshifted 21cm hyperfine transition of intergalactic atomic hydrogen would revolutionise our understanding of cosmological reionization \citep[see e.g. reviews by][]{Furrev,MW10,PL12}. The predicted signal is $\sim 5$ orders of magnitude smaller than the foregrounds, and detecting it presents one of the greatest challenges to modern observational cosmology \citep[e.g.][]{MW10}.

During inhomogeneous reionization Ly$\alpha$ selected galaxies (i.e. LAEs) preferentially reside inside the largest HII regions. LAEs are therefore anti-correlated with the 21cm signal on scales smaller than the characteristic HII bubble size \citep{Lidz09,Wiersma13}. \citet{Lidz09} have quantified the LAE-21cm and `ordinary' (i.e. broad band selected) galaxy-21cm cross power spectra using 130 cMpc cosmological radiative transfer simulations. The amplitude of the LAE-21cm cross power on large scales is higher than that of the galaxy-21cm power spectrum, and should thus be easier to detect\footnote{\citet{Lidz09} note that LAE-21cm cross-correlation is actually sensitive to the characteristic HII regions size around LAEs which are detectable, which especially during the early stages of reionization is larger than the true characteristic HII bubble size. While the LAE-21cm cross-correlation is likely easier detect, it may be more difficult to infer characteristic HII bubble size from this correlation than from the galaxy-21cm correlation.}.

\citet{Lidz09} quantified the significance with which we can detect the LAE-21cm cross-correlation for LOFAR \citep{LOFAR} and MWA\footnote{http://www.mwatelescope.org/}-like 21cm surveys as a function of both $x_{\rm HI}$ and the total field-of-view of a narrowband LAE survey. Their analysis shows that a $2-3\sigma$ detection for a Universe with $x_{\rm HI}\sim 0.5$ requires an LAE field-of-view of $\sim 3$ square degrees, where the significance of the detection increases as the square root of the field-of-view. A detection of this cross-correlation is important is it would confirm that the observed low-frequency signal actually comes from the redshifted 21cm line from the high-redshift IGM.

\vspace{-5mm}
\subsection{Miscalleneous Other Tests}
\vspace{-3mm}
There are several other little explored tests. These include
\vspace{2mm}

{\bf Line Shape Evolution.} There are various processes which can suppress the Ly$\alpha$ flux from galaxies at $z>6$. Each process involves a different radiative transfer effect and is expected to affect the {\it average}\footnote{There exists significant variation in observed line profiles even at a fixed redshift and observed flux, and so we do not expect spectra of in individual galaxies to be able to distinguish between different mechanisms.} line-profile in a different way. Ly$\alpha$ flux suppression by diffuse neutral intergalactic gas and high column density absorbers are expected to leave an imprint of the damping wing on the observed profiles \citep[e.g.][]{Mesinger08}, while models that invoke scattering in the CGM (or evolution in the escape fraction of ionising photons) do not. This possibility has barely been explored quantitatively. \\

{\bf Ly$\alpha$ Halos.} If radiative transfer in the CGM is important - i.e. in case there is a significant flux of photons close to systemic velocity of the galaxy (see Fig~2)- then we expect that the flux that has been removed by the CGM from direct sight lines towards the galaxies to be redistributed over spatially extended diffuse Ly$\alpha$ halos \citep[e.g.][]{Zheng10}.  If the observed reduction in Ly$\alpha$ flux from galaxies at $z>6$ is due to a thickening of the CGM towards higher redshifts, then we would expect there to be a corresponding increase in the total flux in these halos. This is a difficult measurement because the surface brightness of the radiation is very low (and dependent on how far the photons scatter from the galaxy). There are some hints that this has been observed: the observed scale length of Ly$\alpha$ halos appears somewhat larger at $z=6.6$ than at $z=5.7$, but the statistical significance of this effect is weak \citep{Momose14}. Future surveys with Hyper Suprime-Cam should be able to provide better constraints on the redshift evolution of the appearance of Ly$\alpha$ halos. These observations will provide additional constraints on models of Ly$\alpha$ transfer through the ISM/CGM\footnote{In the most extreme case in which outflows shift all photons significantly ($i.e. \gsim 200$ km s$^{-1}$) to the red side of the systemic velocity of the galaxy, we would expect very little scattering in the CGM.}.

\newpage
\section{Conclusions}
\label{sec:conc}

Neutral diffuse intergalactic gas that existed during the Epoch of Reionization (EoR) suppresses Ly$\alpha$ flux emitted by background galaxies. The goal of this chapter was to provide a brief summary of ({\it i}) the intriguing observations of (the lack of) Ly$\alpha$ line emission at $z>6$, ({\it ii}) our current understanding of this observation, ({\it iii}) the challenges we are facing in the modelling, and ({\it iv}) how further surveys will allow us to overcome these challenges. \\

I summarised the increasing observational support for the claim that Ly$\alpha$ photons emitted by galaxies at $z>6$ are suppressed by intervening HI gas in \S~\ref{sec:obs}. I described how this reduction in Ly$\alpha$ flux is manifest in the observed redshift evolution of the Ly$\alpha$ luminosity functions of narrow-band selected galaxies, and in the fraction of broad-band (drop-out) selected galaxies with strong Ly$\alpha$ emission. \\

I described key physical processes that affect Ly$\alpha$ transfer during the EoR in \S~\ref{sec:key}. For clarity, I subdivided the Ly$\alpha$ RT problem into three distinct scales (represented schematically in Fig~\ref{fig:1}): ({\it i}) interstellar scales, which determine the fraction of Ly$\alpha$ photons that manage to escape from the (possibly dusty) ISM of galaxies, but also the spectral line profile of the emerging Ly$\alpha$ flux; ({\it ii}) circumgalactic scales, where Ly$\alpha$ photons can scatter into diffuse halos which may have been observed already. I stressed the dependence of circumgalactic scattering on the emerging Ly$\alpha$ line profile; ({\it iii}) radiative transfer through diffuse neutral intergalactic gas. \\

Ly$\alpha$ transfer during the EoR is clearly a complex, multi scale problem with many uncertainties. I argued in \S~\ref{sec:interpret} that in spite of these uncertainties, the data we have on Ly$\alpha$ emitting galaxies at $z=0-6$ strongly suggests that the observed reduction in Ly$\alpha$ flux from galaxies at $z>6$ is due to additional intervening HI gas. The main question is what fraction of this additional HI gas is in the diffuse neutral IGM. Existing models favor significant contributions from the diffuse IGM, and put constraints on the volume averaged neutral fraction of $x_{\rm HI} \gsim 50\%$. This rapid evolution from $z=6$ is not expected in typical reionization models (though consistent with other observational probes of reionization, including constraints from the Ly$\alpha$ forest and CMB), but it is premature to worry about this as the observational uncertainties are still significant. \\

I finished in \S~\ref{sec:future} with a summary of how future surveys on existing and incoming instruments are expected to reduce existing observational uncertainties enormously by both increasing sample sizes of Ly$\alpha$ emitting galaxies, and extending our sensitivity to much lower Ly$\alpha$ luminosities. With these improved data we will likely be able to nail down reionization with Ly$\alpha$ emitting galaxies.

\begin{acknowledgement}
I would like to thank Andrei Mesinger for permission to reproduce Figures from his work.
\end{acknowledgement}
%
%
%


\end{document}